\RequirePackage{fix-cm}
\documentclass[smallextended]{svjour3}
\smartqed  
\usepackage[table,xcdraw]{xcolor}
\usepackage[utf8]{inputenc}
\usepackage{soul}
\usepackage{graphicx}
\usepackage{nicefrac}
\usepackage{amsfonts}
\usepackage[T1]{fontenc}
\usepackage{xspace}
\usepackage{flushend}
\usepackage{ifthen}
\usepackage{color}
\usepackage{url}
\usepackage{multirow}
\usepackage{listliketab}
\usepackage{amssymb}
\usepackage[linesnumbered,lined,boxed,commentsnumbered]{algorithm2e}
\usepackage{algorithmic}
\usepackage{booktabs}
\usepackage{comment}
\usepackage{amsmath}
\usepackage{bbding}
\usepackage{listings}
\usepackage{float}
\usepackage[misc]{ifsym}
\usepackage{tabularx}
\usepackage{rotating}
\usepackage{natbib}
\usepackage{subfigure}
\usepackage{ulem}
\usepackage{censor}
\usepackage{subcaption}
\usepackage{pgfplots}
\pgfplotsset{compat=1.14}
\usepackage{pgfplotstable}
\usepackage{wrapfig}
\PassOptionsToPackage{hyphens}{url}
\usepackage[hidelinks]{hyperref}
\usepackage{tablefootnote}

\usepackage{microtype} 
\usepackage{threeparttable} 

\definecolor{formalshade}{rgb}{1.0,1.0,1.0}
\definecolor{side}{rgb}{0.0,0.2,0.6}

\definecolor{gray(x11gray)}{rgb}{0.75, 0.75, 0.75}

\lstdefinelanguage{python}{
    morekeywords={access,and,break,class,continue,def,del,elif,else,except,exec,finally,for,from,global,if,import,in,is,lambda,not,or,pass,print,raise,return,try,while},
    morekeywords=[2]{abs,all,any,basestring,bin,bool,bytearray,callable,chr,classmethod,cmp,compile,complex,delattr,dict,dir,divmod,enumerate,eval,execfile,file,filter,float,format,frozenset,getattr,globals,hasattr,hash,help,hex,id,input,int,isinstance,issubclass,iter,len,list,locals,long,map,max,memoryview,min,next,object,oct,open,ord,pow,property,range,raw_input,reduce,reload,repr,reversed,round,set,setattr,slice,sorted,staticmethod,str,sum,super,tuple,type,unichr,unicode,vars,xrange,zip,apply,buffer,coerce,intern},
    sensitive=true,
    morecomment=[l]\#,
    morestring=[b]',
    morestring=[b]",
    morestring=[s]{'''}{'''},
    morestring=[s]{"""}{"""},
    morestring=[s]{r'}{'},
    morestring=[s]{r"}{"},
    morestring=[s]{r'''}{'''},
    morestring=[s]{r"""}{"""},
    morestring=[s]{u'}{'},
    morestring=[s]{u"}{"},
    morestring=[s]{u'''}{'''},
    morestring=[s]{u"""}{"""},
    literate=
    {á}{{\'a}}1 {é}{{\'e}}1 {í}{{\'i}}1 {ó}{{\'o}}1 {ú}{{\'u}}1
    {Á}{{\'A}}1 {É}{{\'E}}1 {Í}{{\'I}}1 {Ó}{{\'O}}1 {Ú}{{\'U}}1
    {à}{{\`a}}1 {è}{{\`e}}1 {ì}{{\`i}}1 {ò}{{\`o}}1 {ù}{{\`u}}1
    {À}{{\`A}}1 {È}{{\'E}}1 {Ì}{{\`I}}1 {Ò}{{\`O}}1 {Ù}{{\`U}}1
    {ä}{{\"a}}1 {ë}{{\"e}}1 {ï}{{\"i}}1 {ö}{{\"o}}1 {ü}{{\"u}}1
    {Ä}{{\"A}}1 {Ë}{{\"E}}1 {Ï}{{\"I}}1 {Ö}{{\"O}}1 {Ü}{{\"U}}1
    {â}{{\^a}}1 {ê}{{\^e}}1 {î}{{\^i}}1 {ô}{{\^o}}1 {û}{{\^u}}1
    {Â}{{\^A}}1 {Ê}{{\^E}}1 {Î}{{\^I}}1 {Ô}{{\^O}}1 {Û}{{\^U}}1
    {œ}{{\oe}}1 {Œ}{{\OE}}1 {æ}{{\ae}}1 {Æ}{{\AE}}1 {ß}{{\ss}}1
    {ç}{{\c c}}1 {Ç}{{\c C}}1 {ø}{{\o}}1 {å}{{\r a}}1 {Å}{{\r A}}1
    {€}{{\EUR}}1 {£}{{\pounds}}1
    {^}{{{\color{ipython_purple}\^{}}}}1
    {=}{{{\color{ipython_purple}=}}}1
    {+}{{{\color{ipython_purple}+}}}1
    {*}{{{\color{ipython_purple}$^\ast$}}}1
    {/}{{{\color{ipython_purple}/}}}1
    {+=}{{{+=}}}1
    {-=}{{{-=}}}1
    {*=}{{{$^\ast$=}}}1
    {/=}{{{/=}}}1,
    literate=
    *{-}{{{\color{ipython_purple}-}}}1
     {?}{{{\color{ipython_purple}?}}}1,
    identifierstyle=\color{black}\ttfamily,
    commentstyle=\color{ipython_cyan}\ttfamily,
    stringstyle=\color{ipython_red}\ttfamily,
    keepspaces=true,
    showspaces=false,
    showstringspaces=false,
    rulecolor=\color{ipython_frame},
    numberstyle=\tiny\color{halfgray},
    backgroundcolor=\color{ipython_bg},
    basicstyle=\scriptsize,
    keywordstyle=\color{ipython_green}\ttfamily,
}

\usepackage[skins,breakable]{tcolorbox}

\definecolor{Large}{HTML}{696969}
\definecolor{Negligible}{HTML}{D3D3D3}
\definecolor{Medium}{HTML}{808080}
\definecolor{Small}{HTML}{A9A9A9}

\newcommand{\best}[1]{{\color{green!60!black}\textbf{#1}}}
\newcommand{\worst}[1]{{\color{red}\textbf{#1}}}

\newcommand{\POne}{{\emph{P1: Multi-repo}\xspace}}
\newcommand{\PTwo}{{\emph{P2: Distribution-only}\xspace}}
\newcommand{\PThree}{{\emph{P3: Designated Dir.}\xspace}}
\newcommand{\PFour}{{\emph{P4: Templating}\xspace}}
\newcommand{\PFive}{{\emph{P5: Bind/Wrap}\xspace}}

\begin{document}

\title{Cross-Ecosystem Packages As Multilingual: Prevalence, Architecture, and Health}

\author{Xiangxi Li \and  Olivier Nourry\and Yoshiki Higo \and  Raula Gaikovina Kula 
}

\date{Received: date / Accepted: date}
\setstcolor{red}
\maketitle

\noindent\textbf{Clinical trial number:} Not applicable.

\begin{abstract}
Modern software development increasingly relies on using multiple programming languages. Some software packages are published to multiple package ecosystems—such as NPM for JavaScript and PyPI for Python.
Little is known about cross-ecosystem packages, especially regarding how they are structured.
In this paper, we conduct a large-scale empirical study of over six million packages across six major ecosystems to understand 1) how prevalent cross-ecosystem packages are among all packages, 2) whether there are distinct source code architectural patterns that cross-ecosystem packages use, and 3) whether there are correlations between architectural patterns and project health metrics from GitHub.
Results indicate that cross-ecosystem packages constitute a small but important, growing fraction of packages.
We identify five distinct architectural patterns. For example, packages that implement code generation from a shared source file or use language bindings are associated with significantly higher community visibility and development activity.
Based on our findings, we provide implications for package adopters, maintainers, and researchers.
We envision our taxonomy being used for future investigations into several aspects of software development, such as the trade-offs between focusing on one language and translating to other languages using bindings, templating, and wrappers, versus using native code and native functions to support additional languages.
\end{abstract}

\section{Introduction}
\label{sec:introduction}

In the context of ecosystems, packages play a unique role in sustaining not only the project itself, but the ecosystem as a whole. 
For software libraries, packages form a supply chain of dependencies, where the failure of a package could lead to disruptions from both downstream and upstream dependents in the ecosystem~\cite{boehmke-gjfsm-2017}.
Furthermore, to counter failures, developers have been encouraged to use traceability measures such as a `software bill of materials (SBOM)' \footnote{\url{https://www.cisa.gov/sbom}} to account for the contents of each package in the supply chain.
However, this might not be as simple for cross-ecosystem packages, as the architectures may differ.
Furthermore, as reported by Yang et al.~\cite{yang-tse-2024}, like any multilingual software, packages may suffer challenges related to \textit{builds, data handling, interoperability, and interfacing (explicit and implicit).}
To the best of our knowledge, no prior work has conducted a large-scale empirical study to systematically characterize the architectural patterns of cross-ecosystem packages. \textbf{Thus, we still do not know how cross-ecosystem packages---which include both libraries and applications---are structured in practice to support developers from multiple ecosystems and facilitate cross-language development}.

In line with related work~\cite{tian-ist-2021}, we define software architecture as the high-level structure of a software system, including the source code. Indeed, Tian et al.~\cite{tian-ist-2021} found that practitioners view software architecture and source code as intertwined artifacts, and that understanding of their relationship is essential for improving maintainability and reliability. Another related work shows that the representation of architecture may take different perspectives of styles, views, patterns, tactics, and decisions~\cite{ali-emse-2018}. In this study, we operationalize architecture at the \textit{repository level}: we focus on observable artifacts such as directory structure, file type distribution, and cross-language integration mechanisms (e.g., language bindings, template-based code generation). This repository-level view is consistent with the SBOM perspective of tracing the concrete source code constituents of a package~\cite{tian-ist-2021}. Hence, in this study, similar to SBOMs, we aim to trace the source code implementations of open-source packages.
This encompasses the file structure, covering both source code and configuration files associated with building packages.
Especially in the case of cross-ecosystem packages, we argue that developers can benefit from knowing whether the packages they use are implemented in their native programming languages or not, as different languages provide different levels of security, reliability, and performance.

In this study, we investigate cross-ecosystem packages across six major ecosystems---Crates.io, Maven Central, NPM, PHP Composer, PyPI, and RubyGems. Our goal is to empirically identify and evaluate the different design strategies used to develop, deploy, and maintain these real-world ecosystem packages. 
Thus, we first collected and compared the packages of the studied ecosystems (i.e., 6,080,775 packages) to determine the prevalence of cross-ecosystem packages. We then investigate how these projects support and deploy source code to multiple ecosystems.
We aim to answer the following two preliminary questions and two research questions:

\begin{itemize}
    \item (PQ1) \textbf{How prevalent are cross-ecosystem packages?} 
    We find that cross-ecosystem packages represent a small but notable fraction of all packages, with thousands of repositories publishing to two or more ecosystems simultaneously.

    \item (PQ2) \textbf{What percentage of cross-ecosystem packages maintain source code in their repositories?} 
    We find that a substantial portion of cross-ecosystem packages lack active source code support in their associated GitHub repositories, indicating that many packages are distributed without a maintained multi-language codebase.
\end{itemize}

Building upon these preliminary results, we further investigate the following research questions:

\begin{itemize}

    \item (RQ1) \textbf{What kinds of source code architectural patterns are employed by cross-ecosystem packages? } We identify five distinct architectural patterns, ranging from loosely coupled multi-repository projects to more tightly integrated binding and wrapper implementations.

    \item (RQ2) \textbf{How do different architectural patterns correlate with project-level health attributes?}
    We find that tighter integration strategies, such as protocol buffers and binding/wrapper projects, are associated with higher community visibility and more active development than loosely structured alternatives.
\end{itemize}

Our results lead us to conclude that cross-ecosystem packages constitute a relatively small but significant proportion of software packages, which has grown since prior research \cite{constantinou-arxiv-2018}. From our investigation, we find that the majority of mono-repo (i.e., a single repository hosting all the code base) cross-ecosystem packages (72.4\%) do not maintain source code for all registered ecosystems, largely due to distribution-only patterns such as Maven WebJars and PHP Composer wrappers. From our analysis of GitHub health metrics and architectural patterns, we also find that maintainers who use tightly integrated strategies (such as wrapper/binding) to achieve cross-ecosystem integration tend to maintain their projects more actively, showing higher commit, contributor, and star counts than loosely structured alternatives.

We make the following contributions:

\begin{enumerate}
    \item \textit{Pattern taxonomy}: We propose a taxonomy of five repository-level architectural patterns for cross-ecosystem packages, derived through an iterative open card sorting process and validated through manual and automated analysis of thousands of repositories.

    \item \textit{Empirical findings}: We provide actionable insights for package adopters, maintainers, dependency management tool developers, and researchers studying software supply chains.

    \item \textit{Replication package}: Our mining scripts, analysis tools, and datasets (package names, homepage URLs, repository URLs, GitHub metrics, directory structures, and source file compositions from six major ecosystems) are publicly available to support the replication and extension of this work.
\end{enumerate}

\section{Background and Related Work}
\label{sec:related_work}

In this section, we define key concepts and then situate our work with respect to the literature.

\subsection{Background}

\textbf{Software ecosystems.} A software ecosystem comprises a package management system along with its associated packages, developer community, and supporting infrastructure~\cite{vandenberk-ecsa-2010}. Major ecosystems include NPM for JavaScript, PyPI for Python, Maven for Java, Crates.io for Rust, Packagist for PHP, and RubyGems for Ruby. Each provides a centralized registry where developers publish packages and declare dependencies.

Prior work has studied the dependency network evolution~\cite{decan-emse-2018}, package popularity~\cite{zerouali-arxiv-2019}, and maintenance practices~\cite{kula-emse-2017, bavota-emse-2015} in these ecosystems. Some work has also documented the dependency vulnerabilities~\cite{prana-emse-2021} and package abandonment~\cite{miller-icse-2025, kula-arxiv-2023} across multiple ecosystems. Furthermore, Valiev et al.~\cite{valiev-fse-2018} found that ecosystem-level factors—such as a project's position in the dependency network—significantly influence sustained project activity. These studies treat each ecosystem independently; our work, instead, focuses on packages that span multiple ecosystems.

\textbf{Cross-ecosystem packages.} A cross-ecosystem package is a software library published to two or more package registries from the same or related source repositories~\cite{constantinou-arxiv-2018}. This is distinct from a simple port or fork: the same project actively maintains multiple language implementations, often under a unified version numbering scheme and release cycle. Prominent examples include Protocol Buffers, gRPC, and major cloud SDKs such as the AWS SDK, each of which is published to many ecosystems in parallel. These libraries bridge language boundaries and serve as foundational infrastructure in multilingual software systems.

\textbf{Multi-language development.} The engineering challenges of maintaining software in multiple programming languages have been investigated through several lenses. Yang et al.~\cite{yang-tse-2024} analyzed Stack Overflow discussions on multi-language programming and identified recurring issues in language interfacing, foreign function calls, and cross-language data handling. Their results show that language boundaries introduce not only syntactic challenges but also coordination overhead in testing and documentation. Maintaining consistent behavior across language implementations requires additional effort that single-language projects do not face. Furthermore, cross-ecosystem packages can be viewed through the lens of system-of-systems architecture~\cite{klein-qosa-2013}, where each language-specific implementation operates independently but is maintained within a larger unified project. Our study complements this perspective by providing large-scale empirical evidence of how cross-ecosystem projects architecturally resolve these challenges in practice.

\subsection{Related Work}

\textbf{Package ecosystem analysis.} Software ecosystems have attracted sustained research attention. Decan et al.~\cite{decan-emse-2018} performed a comparative study of dependency networks across seven package managers, finding substantial variation in dependency growth, breadth, and fragility across ecosystems. Kula et al.~\cite{kula-emse-2017} found that most Java developers do not update their dependencies even when newer versions are available, revealing widespread technical debt in the dependency layer. Abdalkareem et al.~\cite{abdalkareem-fse-2017, abdalkareem-emse-2020} studied the use of trivial packages in NPM and PyPI, finding that developers often depend on packages that implement only minimal functionality. Bogart et al.~\cite{bogart-tosem-2021} examined how 18 open-source ecosystems handle breaking changes, revealing that ecosystems differ substantially in their coordination norms. This difference presents amplified challenges for cross-ecosystem packages that must comply with multiple sets of conventions simultaneously. In this study, we go beyond individual ecosystem analysis to examine packages that are jointly maintained across ecosystems, a dimension that prior ecosystem studies have not addressed.

\textbf{Cross ecosystem package studies.} Constantinou et al.~\cite{constantinou-arxiv-2018} conducted the first large-scale investigation of cross-ecosystem packages by matching repository URLs across 12 package managers, identifying a small but non-negligible set of packages published across ecosystem boundaries. Kannee et al.~\cite{kannee-arxiv-2023} examined the community dynamics of packages spanning multiple ecosystems (NPM, CRAN, Maven, PyPI, and RubyGems) and found that cross-ecosystem packages tend to foster tighter and more interconnected developer communities. Our work extends these studies in three ways: we adopt a consistent unified pipeline across six major ecosystems with a contemporary dataset; we introduce a systematic methodology for detecting whether native source code is actively maintained for each registered ecosystem; and we characterize and compare five distinct repository-level architectural strategies that cross-ecosystem projects employ.

\textbf{Software supply chain security.} The security implications of cross-ecosystem packages are a growing concern. Huang et al.~\cite{huang-emse-2022} characterized the usages, updates, and security risks of third-party libraries in Java projects, finding that outdated dependencies frequently expose projects to security bugs. Wu et al.~\cite{wu-icse-2023} further analyzed upstream vulnerabilities in the Maven ecosystem, showing that a significant portion of downstream projects is affected by vulnerable libraries through transitive dependencies. Williams et al.~\cite{williams-tosem-2025} proposed research directions for software supply chain security, highlighting the role of shared repositories and dependency ecosystems in propagating threats. This body of work motivates our study: because cross-ecosystem packages create implicit inter-ecosystem coupling, understanding their architectural organization is a prerequisite for future security analyses. We note that security is not analyzed in this paper; we return to this as a direction for future work in Section~\ref{sec:discussion}.

\section{Studied Data}
\label{sec:data}

In this section, we describe the data collection pipeline. 
Since our study examines cross-ecosystem packages across six ecosystems, we first need to construct a comprehensive dataset that consists of five stages: 

\begin{enumerate}
    \item Mine package lists from six ecosystems.
    \item Identify cross-ecosystem packages via URL analysis.
    \item Filter out forked, archived, and HTTP-404-error repositories.
    \item Mine directory structures.
    \item Mine GitHub metrics.
\end{enumerate}

\begin{figure}[h]
    \centering
    \includegraphics[width=7cm]{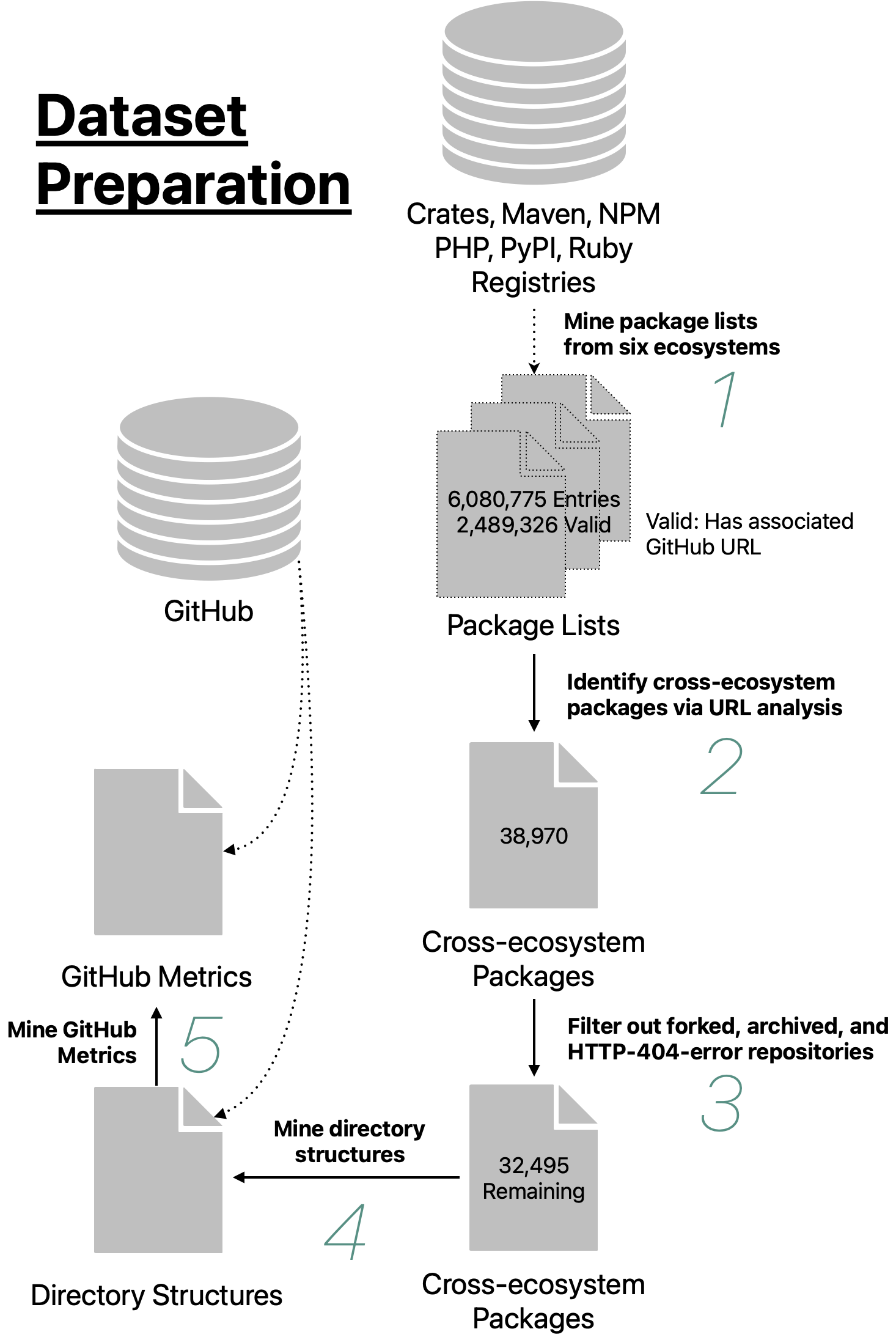}
    \caption{Overview of the dataset preparation}
    \label{fig:dataset}
\end{figure}

Figure~\ref{fig:dataset} shows an overview of our data collection process. We explain how the mined data are used in our preliminary questions (PQs) and research questions (RQs) in the following subsections.

\subsection{Mining Package Lists}
\label{sec:data_mining}

We collect package metadata (Name, Homepage URL, and Repository URL) from six major package ecosystems: Crates.io (Rust)\footnote{\texttt{https://crates.io/}}, Maven Central (Java\-/JVM)\footnote{\texttt{https://central.sonatype.com/}}, NPM (JavaScript/TypeScript)\footnote{\texttt{https://www.npmjs.com/}}, Packagist (PHP)\footnote{\texttt{https://packagist.org/}}, PyPI (Python)\footnote{\texttt{https://pypi.org/}}, and RubyGems (Ruby)\footnote{\texttt{https://rubygems.org/}}.
Data collection was performed in \textbf{[01 2026]};
the dataset reflects a synchronized cross-ecosystem snapshot at that point in time, enabling consistent comparisons across registries.
For each package, we extract the associated GitHub repository URL (either the declared repository URL or the homepage URL) and normalize it to a canonical format\footnote{\texttt{github.com/owner/repo}}.
Across all six ecosystems, we collect 6,080,775 packages linked to 2,489,326 unique normalized GitHub repository URLs.
The difference between the total package count and the unique repository count arises from two sources: (i) many packages across different ecosystems point to the same repository (e.g., a project published to both NPM and PyPI often declares a single shared GitHub URL), and (ii) a subset of packages lack a usable GitHub URL and are excluded from cross-ecosystem detection.

\subsection{Identifying Cross-Ecosystem Packages}
\label{sec:data_identification}

Similar to prior work~\cite{constantinou-arxiv-2018, kannee-arxiv-2023}, we define a \textit{cross-ecosystem package} as a software project that is published to two or more package registries and provides similar or the same service.
This analysis is used as the dataset in PQ1 and PQ2.

\subsection{Filtering Process}
\label{sec:data_filtering}
We filter out repositories that are not accessible, are forks, or are archived. 
Inaccessible repositories are those that return HTTP 404 errors (i.e., repositories that are deleted, renamed, or made private).

\subsection{Mining Directory Structures for Source Code Architecture}
\label{sec:data_directory}

To analyze the source code architecture of the cross-ecosystem packages, we mine the complete directory structures of their repositories using the GitHub API.
The following example shows a sample directory structure of a GitHub repository:

\begin{verbatim}
.github/
.github/workflows/
.github/workflows/release.yml
test/
test/index.js
.gitignore
LICENSE
README.md
index.js
package.json
\end{verbatim}

These directory structures are used in PQ2 and RQ1 to detect the presence of source files (e.g., \texttt{.py} file for PyPI, \texttt{.js} file for NPM, etc.) and analyze the architectural patterns.

\subsection{Mining GitHub Metrics}
\label{sec:data_metrics}

To analyze the relationship between architectural patterns and project health (RQ2), we collect six GitHub metrics for all valid repositories: stars, forks, commits, pull requests, issues, and contributors.
We group these into two categories: \textit{community metrics} and \textit{activity metrics}.
All six metrics represent lifetime totals at the time of data collection; they are not rate-based and do not reflect the pace of development during any particular time window.

\textit{Community metrics}---stars, forks, and contributors---capture community reach and team size~\cite{zerouali-arxiv-2019}.
Specifically, \textbf{stars} reflect community interest and perceived project quality; Borges et al.~\cite{borges-jss-2018} found that three out of four developers consider the number of stars before using or contributing to a GitHub project, making stars a widely used proxy for project adoption and reputation.
\textbf{Forks} indicate how many developers have copied the repository to extend or adapt it, and prior work has shown a moderate-to-strong correlation between forks and stars, reflecting reuse and collaborative intent~\cite{borges-icsme-2016, hu-springerplus-2016}.
\textbf{Contributors} reflect the size of the team that actively commits to the project, capturing the breadth of community involvement.

\textit{Activity metrics}---commits, pull requests, and issues---capture development volume and community engagement.
\textbf{Commits} measure the total volume of code changes and serve as a proxy for overall development activity~\cite{borges-icsme-2016}.
\textbf{Pull requests} reflect active community contribution and collaborative code review, indicating how openly the project accepts external changes.
\textbf{Issues} capture the degree of community interaction through bug reports and feature requests, and are widely used as an indicator of user engagement and project responsiveness~\cite{zerouali-arxiv-2019}.

\section{Preliminary Analysis}
\label{sec:preliminary}

In this section, we present two preliminary analyses that demonstrate how we identify cross-ecosystem packages and detect source code support for them.

\subsection{How Prevalent Are Cross-Ecosystem Packages? (PQ1)}
\label{sec:pq1}

\textbf{Motivation.} Before investigating architectural patterns, we first explore how prevalent cross-ecosystem packages are.
In this analysis, we also deploy different ways to detect cross-ecosystem packages that extend beyond related work. 

\textbf{Approach.}
\label{sec:pq1-approach}
During the initial analysis, we found several patterns in how package URLs are recorded in the registries. 
Hence, we developed two methods to detect cross-ecosystem packages.

\begin{itemize}
    \item \textbf{mono-repo}
This is the classic method used by prior studies\cite{constantinou-arxiv-2018,kannee-arxiv-2023}.
After normalizing all repository URLs, we identify cases where the exact same GitHub URL appears in two or more ecosystem registries.
These packages represent single repositories that are published to multiple ecosystems and are therefore identified as a mono-repo approach.

\item \textbf{multi-repo.}
Initial analysis indicates that some project owners maintain a separate repository for each ecosystem.
We assume that the repositories belong to the same organization on GitHub. 
Some projects maintain separate, language-specific repositories under the same GitHub owner (e.g., \texttt{owner/project-js} and \texttt{owner/project-py}).
Typically, the programming language name is appended as a suffix.
To detect these, we group repositories by owner and strip ecosystem-specific suffixes (e.g., \texttt{-py}, \texttt{-rust}, \texttt{-node}) using curated suffix patterns for each ecosystem.
If two or more repositories from the same owner, in different ecosystems, normalize to the same base name after suffix removal, they are flagged as a multi-repo cross-ecosystem package.
\end{itemize}

\begin{table}[h]
\centering
\begin{threeparttable}
\caption{Package metadata registry entries mined per ecosystem }
\label{tab:package_counts}
\begin{tabular}{lrr}
\toprule
\textbf{Ecosystem} & \textbf{\# URL of Packages} & \textbf{\# Cross-Ecosystem Packages}\\
\midrule
Crates.io & 218,234 & 4,088\\
Maven & 763,405 & 17,004\\
NPM & 3,749,794 & 25,489\\
Packagist (PHP) & 435,167 & 3,975\\
PyPI & 725,184 & 8,908\\
RubyGems & 188,991 & 2,441\\
\midrule
\textbf{Total} & \textbf{6,080,775} & \textbf{61,905}\\
\textbf{Unique Total} & \textbf{2,489,326} & \textbf{38,970}\\
\bottomrule
\end{tabular}
\end{threeparttable}
\end{table}

\textbf{Results.} 
Table~\ref{tab:package_counts} shows the total number of package metadata entries we mined per ecosystem.
The ``Cross-Ecosystem Packages'' column shows how many packages in each ecosystem are identified as cross-ecosystem packages using the two methods mentioned above.
The 38,970 cross-ecosystem packages account for 1.57\% of the 2,489,326 unique GitHub repositories in our dataset.
This number is increasing, as prior work \cite{constantinou-arxiv-2018} reported that 15,389 cross-ecosystem packages accounted for 0.99\% of 1,556,300 packages.

Among these 38,970 packages:

\begin{itemize}
    \item 21,868 (56.1\%) packages use a mono-repo, among which 18,564 are valid (non-forked, non-archived, and non-HTTP-404-error).
    \item 17,102 (43.9\%) packages use a multi-repo, among which 13,930 are valid (non-forked, non-archived, and non-HTTP-404-error).
\end{itemize}

Table~\ref{tab:ecosystem_count} breaks down the ``Cross-Ecosystem Packages'' totals from Table~\ref{tab:package_counts} by the number of ecosystems each package spans (2 through 6), showing how many packages in each ecosystem appear in exactly two, three, four, five, or all six registries simultaneously.
NPM is the most commonly involved ecosystem (25,489 entries), followed by Maven (17,004) and PyPI (8,908).

\begin{table}[t]
\centering
\begin{threeparttable}
\caption{Cross-ecosystem package count by ecosystem and by number of registered ecosystems, with filtering results}
\label{tab:ecosystem_count}
\begin{tabular}{lrrrrrr}
\toprule
\textbf{Ecosystems} & \textbf{2} & \textbf{3} & \textbf{4} & \textbf{5} & \textbf{6} & \textbf{Total} \\
\midrule
Crates   &  3,304 &   500 &  169 &  77 & 38 &  4,088 \\
Maven    & 14,899 & 1,408 &  417 & 242 & 38 & 17,004 \\
NPM      & 22,477 & 2,139 &  584 & 251 & 38 & 25,489 \\
PHP      &  2,343 &   950 &  406 & 238 & 38 &  3,975 \\
PyPI     &  6,412 & 1,701 &  510 & 247 & 38 &  8,908 \\
Ruby     &  1,249 &   520 &  394 & 240 & 38 &  2,441 \\
\midrule
Unique (total)   & 30,119 & 5,135 & 2,221 & 1,286 & 209 &  38,970 \\
\midrule
\multicolumn{6}{l}{\textit{Filtering}} & \\
\multicolumn{6}{l}{\quad HTTP-404 errors\tnote{a}} & 2,816 \\
\multicolumn{6}{l}{\quad Forked repositories\tnote{b}} & 731 \\
\multicolumn{6}{l}{\quad Archived repositories\tnote{c}} & 3,000 \\
\multicolumn{6}{l}{\quad Total Removed: Forked $\cup$ Archived $\cup$ Errors} & 6,475 \\
\midrule
\multicolumn{6}{l}{\textbf{Valid total}} & \textbf{32,494} \\
\multicolumn{6}{l}{\quad Mono-repo (URL-matched)} & 18,564 \\
\multicolumn{6}{l}{\quad Multi-repo (URL-not-matched)} & 13,930 \\
\bottomrule
\end{tabular}
\begin{tablenotes}
\footnotesize
\item[a] Repositories deleted, renamed, or made private (HTTP 404 via GitHub API).
\item[b] Forked repositories excluded; 71 repositories are counted once here as both forked and archived.
\item[c] Archived repositories excluded from further analysis.
\end{tablenotes}
\end{threeparttable}
\end{table}

\begin{tcolorbox}
Cross-ecosystem packages are not prevalent. However, they still represent a growing number when compared to prior work.
\begin{itemize}
    \item Observation 1: Cross-ecosystem packages may span multiple GitHub repositories.
\end{itemize}

\end{tcolorbox}

\subsection{Do Cross-Ecosystem Packages Maintain Source Code for All Registered Ecosystems? (PQ2)}
\label{sec:pq2}

\textbf{Motivation.} Being registered in multiple ecosystems does not guarantee that a repository actually maintains source code for all of them. Before studying architectural patterns in depth (RQ1), we investigate whether cross-ecosystem packages actively maintain source code support for their registered ecosystems by checking if source files appear in their directory structures. 

\textbf{Approach.}
Table~\ref{tab:ecosystem_count} shows the filtering process applied to obtain valid repositories for answering PQ2. 
We use non-forked, non-archived, and non-HTTP-404-error packages in the following analysis, yielding:

\begin{itemize}
    \item 18,564 mono-repo packages.
    \item 13,930 multi-repo packages.
\end{itemize}

We conduct two analyses based on whether packages use multi-repo or mono-repo structures.
For 13,930 multi-repo packages, each repository is expected to serve a single ecosystem as indicated by the language-specific suffix in its name (e.g., \texttt{repo-py} $\rightarrow$ Python).
To verify this assumption, we check whether the expected programming language actually appears in each repository's GitHub-reported languages.

For each of the 18,564 mono-repo packages, we scan their GitHub repository directory structure for source file extensions associated with each studied ecosystem (using the mined directory structures).
We iterate over every file path in the mined directory structure and extract each file's extension. Then, we look up the extracted file extension against a predefined extension-to-ecosystem mapping to determine which ecosystem(s) are present. 
Each ecosystem is identified by its characteristic source file types:

\begin{itemize}
    \item PyPI maps to Python files (\texttt{.py, .pyx, .pxd, .pyi}), 
    \item Crates to Rust files (\texttt{.rs}),
    \item NPM to JavaScript and TypeScript files (\texttt{.js, .jsx, .ts, .tsx, .mjs, .cjs, .css, .scss}), 
    \item Maven to JVM-based files (\texttt{.java, .scala, .kotlin, .kt}), 
    \item Ruby to Ruby files (\texttt{.rb, .rake}),
    \item   and PHP to PHP files (\texttt{.php}).
\end{itemize}

We exclude common non-source directories (e.g., test, documentation, build, vendor, and cache folders) using approximately 40 exclusion keywords to avoid false detections.
A single source file is sufficient to consider an ecosystem as ``detected.''

\begin{table}[t]
\centering
\begin{threeparttable}
\caption{Multi-repo language verification: expected language presence by ecosystem}
\label{tab:multi-repo_lang_verification}
\begin{tabular}{lrrr}
\toprule
\textbf{Ecosystem} & \textbf{Matched}\tnote{1} & \textbf{Mismatched}\tnote{2} & \textbf{Match Rate}\tnote{3} \\
\midrule
Crates &    694 &   5 & 99.28\% \\
Maven  &  1,975 &  30 & 98.50\% \\
NPM    &  4,217 &  70 & 98.37\% \\
PHP    &  2,482 &  27 & 98.92\% \\
PyPI   &  3,168 &  24 & 99.25\% \\
Ruby   &  1,231 &   9 & 99.27\% \\
\midrule
\textbf{Total} & \textbf{13,765} & \textbf{165} & \textbf{98.82\%} \\
\bottomrule
\end{tabular}
\begin{tablenotes}
\footnotesize
\item[1] Matched: Expected language appeared in the GitHub repository's language proportions. (e.g. \texttt{github.com/owner/repo-py} has Python)
\item[2] Mismatched: Expected language doesn't appear in the GitHub repository's language proportions. (e.g. \texttt{github.com/owner/repo-py} only has JavaScript)
\item[3] Match Rate: The proportion of matched packages among all packages in its corresponding ecosystem.
\end{tablenotes}
\end{threeparttable}
\end{table}

\textbf{Results.} 
We divide the results based on the multi-repository and mono-repository analysis. 

\textit{Multi-repo packages.}
Table~\ref{tab:multi-repo_lang_verification} shows the per-ecosystem breakdown.
Of the valid multi-repo repositories, a majority (98.82\%) have their expected language present in the repository's GitHub-reported languages, as expected.

All evaluated ecosystems exhibit match rates exceeding 98\%, with Crates, PyPI, and Ruby achieving the highest scores. The remaining 165 repositories exhibit mismatches attributable to three primary causes: (1) empty repositories yielding no language metadata from GitHub; (2) suffixes that coincidentally match a language token but semantically represent non-language artifacts---e.g., the ``lang-java'' package, which is a CodeMirror language support module implemented in TypeScript; and (3) repositories that have been repurposed for alternative use cases, thereby diverging from their original language designation.

\textit{Mono-repo packages}.
We classify the 18,564 mono-repo packages into three categories based on how well their detected ecosystems match their registered ecosystems:
\begin{itemize}
    \item Fully Matched: 5,132 packages (27.6\%) --- all registered ecosystems are detected in the repository.
    For example, \textit{github.com/kreuzberg-dev/html-to-markdown}\footnote{\url{https://github.com/kreuzberg-dev/html-to-markdown}} is registered in Crates, NPM, PHP, PyPI, and Ruby ecosystems.
    All source files for all five different ecosystems are found in this single repository.
    \item Partially Matched: 171 packages (0.9\%) --- two or more (but not all) registered ecosystems are detected.
    For example, \textit{github.com/asimov-modules/asimov-apify-module}\footnote{\url{https://github.com/asimov-modules/asimov-apify-module}} is registered in Crates, NPM, PyPI, and Ruby, but only Crates and Ruby source files are found.
    \item Fully Mismatched: 13,261 packages (71.4\%) --- one or zero ecosystem source files are detected.
    For example, \textit{github.com/orbitinghail/graft}\footnote{\url{https://github.com/orbitinghail/graft}} is registered in Crates, NPM, PyPI, and Ruby, but only Crates source files are found, with no evidence of NPM, PyPI, or Ruby source code.
\end{itemize}

The majority of mono-repo cross-ecosystem packages do not maintain native source code for all their registered ecosystems within their repository.

\begin{tcolorbox}
Most multi-repo cross-ecosystem packages store a single source code language, while a majority of mono-repo cross-ecosystem packages do not maintain source files for every one of their registered ecosystems.
\begin{itemize}
    \item Observation 2: A majority of mono-repo packages (i.e., 72.4\%) do not maintain native source code for all of their registered ecosystems within their repository (combining partially matched 0.9\% and fully mismatched 71.4\%).
\end{itemize}
\end{tcolorbox}

\section{What Architectural Patterns Do Cross-Ecosystem Packages Employ? (RQ1)}
\label{sec:rq1}
Based on the preliminary results, we move to the empirical study. In this section, we present how we define architectural patterns and how we detect them in our dataset.

\textbf{Motivation.} Having established the prevalence of cross-ecosystem packages (PQ1) and the extent of source code presence (PQ2), we now investigate what architectural patterns these packages employ---including the causes of the high mismatch rate observed in PQ2. Understanding these patterns enables tool builders and researchers to better support cross-ecosystem development.

\textbf{Approach.}
To derive the architectural patterns, we followed an iterative open card sorting process~\cite{spencer-rosenfeld-2009} inspired by prior work on repository classification.
The process was as follows:

\begin{enumerate}
    \item We drew a stratified random sample of 50 repository URLs from the two subsets established in the preliminary analysis (mismatched mono-repo packages and fully matched mono-repo packages).
    \item Each author independently inspected each repository and assigned a short descriptive label capturing the repository's cross-ecosystem organization strategy.
    \item The first author held meetings with the other two authors to compare labels, resolve disagreements, and merge semantically equivalent labels into candidate patterns. In total, 12 such meetings were conducted throughout the process.
    \item If any new pattern was identified in this round, a fresh sample of 50 URLs was drawn and the process repeated from step~2.
    \item The process stopped when a full round of 50 samples produced no new patterns.
\end{enumerate}

After 41 iterations spanning approximately 12 weeks, no further new patterns emerged, yielding the five patterns presented below.
Note that \POne~(multi-repo) was identified during PQ1 and was therefore not part of the card sorting sample pool; the card sorting focused exclusively on mono-repo packages, stratified into mismatched and fully matched groups.
\PTwo~(distribution-only) emerged from the targeted analysis of PQ2 mismatches and was subsequently confirmed across card sorting iterations.
The automated detection rules for each pattern were then designed and validated against the samples accumulated during card sorting.
We describe each pattern's definition and detection rule, and how they relate to the datasets established in prior sections.

For the analysis, we tally all the different patterns and perform a detailed analysis of each. Note that we use semi-automatic detection based on the heuristics derived during the card sorting process. 
It is important to note that these classifications are not mutually exclusive; a package can exhibit more than one pattern.

\begin{table}[h]
\centering
\caption{Pattern identification results}
\label{tab:pattern_results}
\begin{tabular}{p{4cm}rr}
\toprule
\textbf{Pattern} & \textbf{GitHub owners} & \textbf{Unique Repos} \\
\midrule
\POne          &  5,816 & 13,930 \\
\PTwo     & 13,432  & 13,432 \\
\PThree  &  2,233 &  2,233 \\
\PFour       &    474  &    474 \\
\PFive           &  1,814 &  1,814 \\
\bottomrule
\end{tabular}
\end{table}

\textbf{Results.}
Table~\ref{tab:pattern_results} summarizes the five patterns that we identified in our card sorting process.
Note that for \POne, the ``GitHub owners" is much less than its ``Unique Repos" because we group the repositories that belong to the same project by owner.
Additionally, we find 1,566 repositories that appear in multiple patterns.
The most common overlap is between \PThree~and \PFive, with 732 shared repositories.
The second most common overlap is between \PTwo~and \PFive, with 537 shared repositories.
We now discuss each pattern in detail. 

\paragraph{\textbf{\POne~(from Observation 1.)}}
The first pattern is taken from the analysis of PQ1.
In this pattern, cross-ecosystem packages have related repositories that share the same GitHub owner but use distinct URLs with language-specific suffixes.
Each ecosystem is served by a separate repository rather than a single shared codebase.
The 5,816 multi-repository groups span 13,930 individual repositories, meaning each group contains, on average, 2.4 repositories.
NPM (4,149) and PHP (2,506) account for the largest shares, while Crates (694) is the least common.

We describe an example of this pattern using the OpenAI SDK package.
This package is managed through different URLs but is owned by OpenAI. The naming convention appends the programming language ecosystem as a suffix\footnote{ An example for Java is openai-java. \url{github.com/openai/openai-java}}.
The package is also registered for NPM (i.e., openai-node), Python (i.e., openai-python), and RubyGems (i.e., openai-ruby).

\paragraph{\textbf{\PTwo~(from Observation 2.)}}
The second pattern was detected during PQ2 and is particularly prevalent among mono-repo packages.
Rather than a methodology limitation, this represents a deliberate \textit{distribution strategy}: the repository intentionally publishes to additional ecosystems via pre-built or repackaged artifacts, without maintaining native source code for those ecosystems.

To characterize this pattern, we performed a targeted analysis and manual inspection of 200 sampled mismatched packages, identifying two dominant sub-patterns.
The first is Maven WebJar/mvnpm publishing: we found that 11,083 (59.7\% of all mono-repo packages) were actually WebJar or mvnpm packages (with \texttt{groupId} starting with \texttt{org.webjars.*} or \texttt{org.mvnpm.*}).
These packages repackage front-end JavaScript assets for consumption via Maven, without including Java source code.
The second sub-pattern involves PHP Composer wrappers: we find 499 (2.7\%) packages that contain a \texttt{composer.json} file (distributing front-end assets to PHP via Composer) but no actual \texttt{.php} source files.

An example of this distribution-only strategy is the vue package\footnote{\url{github.com/vuejs/vue}}.
This package is published to both NPM and Maven, but the repository contains only JavaScript; the Maven publication is a WebJar repackaging of the JavaScript build artifact, not a separate Java implementation.

\paragraph{\textbf{\PThree}}
The third pattern is based on the directory structure within the repository.
During the card sorting, we found that some cross-ecosystem packages designated a specific folder for each target ecosystem. 
Hence, we searched for directories matching the following language-specific names.
\begin{itemize}
    \item  PyPI $\rightarrow$ \texttt{python}, \texttt{py}, etc.;
    \item NPM $\rightarrow$ \texttt{js}, \texttt{javascript}, etc.; 
    \item Crates $\rightarrow$ \texttt{rust}, \texttt{rs}, etc.; 
    \item Maven $\rightarrow$ \texttt{java}, \texttt{jvm}, etc.; 
    \item Ruby $\rightarrow$ \texttt{ruby}, \texttt{rb}, etc.; 
    \item PHP $\rightarrow$ \texttt{php}, etc.
\end{itemize}

A package is classified under this pattern if two or more language-specific folders are found.
To validate the folders, we additionally compute the \textit{coverage ratio} (the proportion of that ecosystem's source files inside the folder) and classify packages as:
\begin{itemize}
    \item Concentrated-Complete: All ecosystem folders found with each $\geq 80\%$ coverage.
    \item Concentrated-Partial: Some ecosystem folders were found with each $\geq 80\%$ coverage.
    \item Mixed: Some folders were $\geq 80\%$, while others were $\geq 40\%$ coverage.
    \item Low: All found folders had $< 40\%$ coverage.
\end{itemize}

\begin{table}[h]
\centering
\caption{Directory structure classification of all fully matched mono-repo packages ($n = 5{,}132$); only packages in the top four rows (with language-specific folders) constitute \PThree~($n = 2{,}233$)}
\label{tab:naming_convention}
\begin{tabular}{lr}
\toprule
\textbf{Classification} & \textbf{Count (\%)} \\
\midrule
Concentrated-Complete &    659 (12.8\%) \\
Concentrated-Partial  &     90 (1.8\%) \\
Mixed                 &  1,272 (24.8\%) \\
Low                   &    212 (4.1\%) \\
No Designated Directory & 2,899 (56.5\%) \\
\bottomrule
\end{tabular}
\end{table}

Table~\ref{tab:naming_convention} presents the directory structure classification applied to all 5,132 fully matched mono-repo packages.
Only the packages in the top four categories---those with at least one language-specific folder---are classified as \PThree~(totaling 2,233 packages).
The ``No Designated Directory'' row (2,899 packages, 56.5\%) represents fully matched packages whose source files are \textit{not} organized into named ecosystem folders; these packages are \textit{not} classified as \PThree.
Among \PThree~packages, 659 (12.8\% of the fully matched set) achieve a Concentrated-Complete structure, where every ecosystem's source files are concentrated ($\geq 80\%$) within its named folder.

One example of this pattern is the react-native package \footnote{\url{github.com/facebook/react-native}}.
In this example, the package stores the Java/Kotlin code under the directory ReactAndroid/src/main/java/, while the JavaScript is placed in a separate directory at \texttt{flow-typed/npm/}. 
It has a coverage of 92.7\% of the detected matching source files in those directories. 

\paragraph{\textbf{\PFour}}
The fourth pattern relates to mechanisms used to generate code in multiple languages. 
Hence, we explored different technologies used for templating. 
Specifically, packages using interface definition languages (IDLs) or template-based code generation produce language-specific bindings from shared schema definitions.
Based on our card sorting, we identified three kinds of templating (i.e., \texttt{.proto} (Protocol Buffers), \texttt{.thrift} (Apache Thrift), and \texttt{.fbs} (FlatBuffers)).

As shown in Table \ref{tab:naming_convention}, we find that 474 packages use IDL-based code generation, with NPM (319) and PyPI (254) being the most common target ecosystems; PHP appears in only 4 packages.

An example of such package is flatbuffers \footnote{\url{github.com/google/flatbuffers}}. 
In this case, the package utilizes the \texttt{.fbs} schema files to generate templates for Crates, Maven, NPM, PHP, and PyPI simultaneously.

\paragraph{\textbf{\PFive}}
The final pattern is similar to P4, but instead of templates, the mechanism is language-binding technologies.
Similar to P2, we also found cases where pre-built binary distributions were published. 
For P5, we identified the following strategies:

\begin{itemize}
    \item Specific Bindings---detect JSII, WASM, and PyO3/Maturin bindings.
     
    \item General Bindings---detect binding directories or files\footnote{Examples include looking for the following keywords: \texttt{binding/}, \texttt{ffi/}, \texttt{napi/}, \texttt{jni/}, \texttt{cgo/}, \texttt{pybind/}, \texttt{cython/}, \texttt{binding.gyp}, \texttt{.node}, \texttt{.pyd}}.
    
    \item OS Distributions---detect operating system platform directories used for distributing binaries as wrappers\footnote{
    Folders in the repository match combined OS-architectural patterns (e.g., \texttt{linux-x86\_64/}, \texttt{darwin-arm64/}, \texttt{windows-amd64/}), indicating prebuilt native binaries distributed as wrappers}.
\end{itemize}

Table~\ref{tab:binding_sources} presents the breakdown by detection source. Specific bindings are the most prevalent, with 1,143 detected packages. Of these, 668 use WASM bindings, 405 use JSII bindings, and 70 use PyO3/Maturin. Besides these, we also detect 735 packages with general binding indicators, followed by 282 OS distribution packages. Across all sources, we record 2,160 detections; after removing 346 duplicates, 1,814 unique packages exhibit binding or wrapper patterns.

An example of the final pattern is the next.js package \footnote{\url{github.com/vercel/next.js}}. 
In this example, the package employs WASM modules to compile Rust code for the Crates ecosystem.

\begin{table}[t]
\centering
\caption{\PFive~detection sources}
\label{tab:binding_sources}
\begin{tabular}{lr}
\toprule
\textbf{Source} & \textbf{\# packages} \\
\midrule
Specific Bindings   & 1,143 \\
General Bindings    & 735 \\
OS Distributions   & 282 \\
\midrule
Total (before dedup) & 2,160 \\
Duplicates   &   346 \\
\textbf{Total (after dedup)} & \textbf{1,814} \\
\bottomrule
\end{tabular}
\end{table}

\begin{tcolorbox}
We identify five distinct repository-level architectural patterns. Most cross-ecosystem packages do not maintain native source code for all registered ecosystems within their repository; many instead follow a distribution-only strategy.
\begin{itemize}
    \item Observation 3. Cross-ecosystem packages employ different cross-language mechanisms---templating, bindings, or wrappers---to publish to multiple ecosystems without duplicating native source code.
    \item Observation 4. Some cross-ecosystem packages organize their repository with language-specific designated directories, co-locating each ecosystem's source code within a dedicated folder.
\end{itemize}
\end{tcolorbox}

\section{How Do Architectural Patterns Correlate with Project Health? (RQ2)}
\label{sec:rq2}

In this section, we investigate how the architectural patterns identified in RQ1 are associated with project health, which we characterize using six project metrics shown in Table~\ref{tab:descriptive_stats}.

\textbf{Motivation.} Having identified five architectural patterns (RQ1), we now investigate whether different patterns are associated with different project health and community profiles. This analysis helps practitioners understand the observable differences between organizational strategies for cross-ecosystem development.

\begin{table}[h]
\centering
\caption{Descriptive statistics of GitHub metrics per pattern}
\label{tab:descriptive_stats}
\begin{threeparttable}
\begin{tabular}{llrrr}
\toprule
\textbf{Metric} & \textbf{Pattern} & $n$ & \textbf{Median} & \textbf{Mean} \\
\midrule
\multirow{5}{*}{Stars}
 & \POne       & 5,816 & \worst{6} & \worst{221.6} \\
 & \PTwo  & 13,432 & \best{71} &  2,015.0 \\
 & \PThree  & 2,233 &   19 &  1,585.1 \\
 & \PFour      &   474 &   68 &  \best{3,750.1} \\
 & \PFive       & 1,814 &   22 &  2,486.6 \\
\midrule
\multirow{5}{*}{Forks}
 & \POne       & 5,816 & \worst{3} &  \worst{65.2} \\
 & \PTwo  & 13,432 & 16 &   291.2 \\
 & \PThree  & 2,233 &  5 &   292.9 \\
 & \PFour      &   474 & \best{20} &   \best{754.9} \\
 & \PFive       & 1,814 &  4 &   396.4 \\
\midrule
\multirow{5}{*}{Commits}
 & \POne       & 5,816 & 112 &  \worst{663.8} \\
 & \PTwo  & 13,432 & \worst{107} &    721.3 \\
 & \PThree  & 2,233 & 250 &  2,083.1 \\
 & \PFour      &   474 & \best{580} &  \best{4,751.2} \\
 & \PFive      & 1,814 & 300 &  3,170.1 \\
\midrule
\multirow{5}{*}{Pull Requests}
 & \POne       & 5,816 & \worst{14} &  \worst{268.6} \\
 & \PTwo  & 13,432 &  23 &   281.4 \\
 & \PThree  & 2,233 &  53 &   922.1 \\
 & \PFour      &   474 & \best{178} & \best{2,395.7} \\
 & \PFive       & 1,814 &  67 &  1,141.0 \\
\midrule
\multirow{5}{*}{Issues}
 & \POne       & 5,816 & \worst{2} &  \worst{89.9} \\
 & \PTwo  & 13,432 & 13 &   276.5 \\
 & \PThree  & 2,233 & 11 &   462.0 \\
 & \PFour      &   474 & \best{37} & \best{1,019.5} \\
 & \PFive       & 1,814 &  7 &   579.4 \\
\midrule
\multirow{5}{*}{Contributors}
 & \POne       & 5,816 &  5 &  \worst{25.1} \\
 & \PTwo  & 13,432 &  7 &   37.0 \\
 & \PThree  & 2,233 &  6 &   53.4 \\
 & \PFour      &   474 & \best{17} &  \best{131.3} \\
 & \PFive       & 1,814 & \worst{4} &   82.2 \\
\bottomrule
\end{tabular}
\end{threeparttable}
\end{table}

\begin{table}[h]
\centering
\caption{Pattern distribution: Top 10\% vs.\ Bottom 10\% by community metrics ($n = 2{,}376$ each)}
\label{tab:top_bottom_popularity}
\begin{tabular}{lrr}
\toprule
\textbf{Pattern} & \textbf{Top 10\%} & \textbf{Bottom 10\%} \\
\midrule
\multicolumn{3}{c}{\textbf{Stars}} \\
\midrule
\POne     & \worst{91 ( 3.8\%)}   & \best{1,112 (46.8\%)} \\
\PTwo     & \best{1,742 (73.3\%)}  & 776 (32.7\%)          \\
\PThree   & 220 ( 9.3\%)           & 239 (10.1\%)          \\
\PFour    & 98 ( 4.1\%)            & \worst{52 ( 2.2\%)}   \\
\PFive    & 225 ( 9.5\%)           & 197 ( 8.3\%)          \\
\midrule
\multicolumn{3}{c}{\textbf{Forks}} \\
\midrule
\POne     & \worst{195 ( 8.2\%)}  & \best{937 (39.4\%)}   \\
\PTwo     & \best{1,651 (69.5\%)} & 867 (36.5\%)          \\
\PThree   & 224 ( 9.4\%)           & 287 (12.1\%)          \\
\PFour    & 102 ( 4.3\%)           & \worst{45 ( 1.9\%)}   \\
\PFive    & 204 ( 8.6\%)           & 240 (10.1\%)          \\
\midrule
\multicolumn{3}{c}{\textbf{Contributors}} \\
\midrule
\POne     & 414 (17.4\%)           & 662 (27.9\%)          \\
\PTwo     & \best{1,327 (55.9\%)}  & \best{1,220 (51.3\%)} \\
\PThree   & 268 (11.3\%)           & 236 ( 9.9\%)          \\
\PFour    & \worst{119 ( 5.0\%)}   & \worst{32 ( 1.3\%)}   \\
\PFive    & 248 (10.4\%)           & 226 ( 9.5\%)          \\
\bottomrule
\end{tabular}
\end{table}

\begin{table}[h]
\centering
\caption{Pattern distribution: Top 10\% vs.\ Bottom 10\% by activity metrics ($n = 2{,}376$ each)}
\label{tab:top_bottom_activity}
\begin{tabular}{lrr}
\toprule
\textbf{Pattern} & \textbf{Top 10\%} & \textbf{Bottom 10\%} \\
\midrule
\multicolumn{3}{c}{\textbf{Commits}} \\
\midrule
\POne     & 421 (17.7\%)           & 646 (27.2\%)          \\
\PTwo     & \best{1,032 (43.4\%)}  & \best{1,482 (62.4\%)} \\
\PThree   & 398 (16.8\%)           & 135 ( 5.7\%)          \\
\PFour    & \worst{159 ( 6.7\%)}   & \worst{16 ( 0.7\%)}   \\
\PFive    & 366 (15.4\%)           & 97 ( 4.1\%)           \\
\midrule
\multicolumn{3}{c}{\textbf{Pull Requests}} \\
\midrule
\POne     & 397 (16.7\%)           & \best{927 (39.0\%)}   \\
\PTwo     & \best{1,024 (43.1\%)}  & 1,089 (45.8\%)        \\
\PThree   & 408 (17.2\%)           & 207 ( 8.7\%)          \\
\PFour    & \worst{156 ( 6.6\%)}   & \worst{26 ( 1.1\%)}   \\
\PFive    & 391 (16.5\%)           & 127 ( 5.3\%)          \\
\midrule
\multicolumn{3}{c}{\textbf{Issues}} \\
\midrule
\POne     & 252 (10.6\%)           & \best{942 (39.6\%)}   \\
\PTwo     & \best{1,438 (60.5\%)}  & 949 (39.9\%)          \\
\PThree   & 297 (12.5\%)           & 224 ( 9.4\%)          \\
\PFour    & \worst{122 ( 5.1\%)}   & \worst{44 ( 1.9\%)}   \\
\PFive    & 267 (11.2\%)           & 217 ( 9.1\%)          \\
\bottomrule
\end{tabular}
\end{table}

\textbf{Approach.}
Using the project metrics we collected from GitHub, we conduct two analyses.
First, for each pattern found in RQ1, we compute descriptive statistics (median, mean) for each project health metric.

For the second analysis, we independently rank all cross-ecosystem packages across the six metrics to conduct a stratified analysis of the top and bottom deciles. By calculating the frequency of each pattern within these highest and lowest 10\% thresholds, we measure the over- or under-representation of each pattern at the metric extremes.
Ties at the 10\% boundary are broken by random sampling with a fixed seed for reproducibility.

\textbf{Results.}

\textit{Note on interpretation.} All results in this section report statistical associations between architectural patterns and project health metrics. Because this is an observational study, we do not claim a causal relationship; the patterns may co-occur with certain health profiles due to confounding factors (e.g., project age, domain, or team size) that we do not control for.

\begin{sloppypar}

For the first analysis (Table~\ref{tab:descriptive_stats}), we find that
\textbf{\PTwo~packages are associated with the highest star counts, while \PFour~packages show the highest development activity across forks, commits, pull requests, issues, and contributors.}

Regarding community metrics, \PTwo~packages have the highest median star count (71), followed by \PFour~(68). \PTwo~packages also have the second-highest median forks (16), just behind \PFour~(20). This association suggests that distribution-only packages---such as JavaScript libraries repackaged as Maven WebJars---are widely recognized in the community. Interestingly, this high visibility occurs even though they lack native source code for some ecosystems, which may indicate that users are unaware of the absence of a native implementation.
    
\end{sloppypar}

Despite their popularity, \PTwo~packages show lower development activity. Their median commits (107) are similar to \POne~(112) but much lower than \PFour~(580) and \PFive~(300). Similarly, their median pull requests (23) are far below \PFour~(178), \PFive~(67), and \PThree~(53). This combination of high popularity and low activity is common when a package's reputation is tied to the original source-language project, rather than active cross-ecosystem development.

In contrast, \PFour~packages have the highest median values across the board for activity: commits (580), pull requests (178), issues (37), and contributors (17). These higher metrics frequently correspond with template-generation projects, which often serve infrastructure needs and involve larger teams.

Finally, \POne~has the lowest median values for all metrics (e.g., 6 stars, 3 forks, and 5 contributors). We observe that these lower numbers consistently accompany projects that split their development across separate, per-language repositories.

\begin{sloppypar}

For the second analysis (Tables~\ref{tab:top_bottom_popularity} and~\ref{tab:top_bottom_activity}), we find that
\textbf{\PTwo~is over-represented in the top tier for community metrics but under-represented for activity, whereas \PThree, \PFour, and \PFive~skew toward higher activity.}
    
\end{sloppypar}

For popularity metrics, \PTwo~is concentrated in the top 10\%, accounting for 73.3\% of the most-starred packages, 69.5\% of the most-forked, and 55.9\% of those with the most contributors. However, for activity metrics, \PTwo~are toward the bottom: 62.4\% of the least-committed packages are \PTwo, compared to 43.4\% in the top 10\%. 

\POne~constitutes 46.8\% of the bottom 10\% by stars but only 3.8\% of the top, and 39.0\% of the bottom by pull requests but only 16.7\% of the top tier.
Conversely, tightly integrated patterns (\PThree, \PFour, \PFive) consistently skew toward higher activity. \PFour~appears at 6.7\% of the top 10\% by commits but only 0.7\% of the bottom---a ratio of roughly $10\times$. \PFive~and \PThree~show similar skews, with top-10\% representation exceeding their bottom-10\% share by approximately 2--3$\times$ across activity metrics. 

\begin{tcolorbox}
We find that different architectural patterns are associated with different health metric profiles. Distribution-only packages (\PTwo) are among the most visible cross-ecosystem packages by stars and forks, yet they exhibit relatively low development activity.
\begin{itemize}
    \item Observation 5. \PTwo~packages are associated with the highest star counts, while \PFour~packages show the highest development activity (forks, commits, pull requests, issues, and contributors).
    \item Observation 6. \PTwo~is over-represented in the top decile for community metrics but under-represented for activity metrics, whereas \PThree, \PFour, and \PFive~skew toward higher activity.
\end{itemize}
\end{tcolorbox}

\section{Discussion}
\label{sec:discussion}

In this section, we discuss the implications for package adopters, package maintainers, researchers, and tool builders. In addition to these Implications, we also highlight different areas for future investigation.

We find that most mono-repo cross-ecosystem packages do not maintain native source code for all registered ecosystems within their repository (Observation 2); the majority follow a distribution-only strategy where one language's artifacts are repackaged for other ecosystems. At the same time, packages using templating or binding strategies are among the most actively developed (Observations 5 and 6). Developers should therefore be aware that a package imported from another ecosystem may not have a native implementation maintained for their target language. Our taxonomy directly supports this judgment: a package classified as \PTwo~(distribution-only) versus \PFour~(templating) or \PFive~(binding) carries meaningfully different implications for long-term language-specific support and maintenance. Based on these findings, developers can now make a more informed decision about whether a cross-ecosystem package suits their supply chain needs or whether a native alternative is preferable. Future developer surveys could elicit how aware adopters are of these distribution strategies and whether they affect adoption decisions.

\subsection{Implications for package maintainers.}

A key finding of this research is that several strategies can be effective to develop these multilingual packages (Observation 1, Observation 2, Observation 3, Observation 4). Specifically, we found that there is a significant number of packages that are distribution only (\PTwo~and Observation 2). For package maintainers currently working on supporting several languages natively, our results show that a significant amount of cross-ecosystem packages are able to successfully deploy to several ecosystems while maintaining a single programming language. Our results provide a good overview of the field and allow package maintainers to consider the use of tools to support additional languages.

\subsection{Implications for Researchers and Tool Builders.}
For researchers, our investigation provides fine-grained (directory-level) empirical evidence that cross-ecosystem packages can be implemented and deployed in several ways. This study provides an operational taxonomy for conducting further research on open-source communities, development practices, and software ecosystems. The taxonomy is grounded in detectable repository signals and supported by manual validation, making it directly applicable to automated classification.

For tool builders, our taxonomy has concrete applications in dependency analysis and SBOM generation. SBOM tools could use the architectural pattern of a package to tag it as \textit{native}, \textit{templated}, or \textit{bound/wrapped}, enriching supply chain metadata beyond simple package name and version. For example, knowing that a Maven package is a \PTwo~(distribution-only) WebJar repackaging of a JavaScript library---rather than a native Java implementation---is directly relevant for vulnerability assessment, license compliance, and maintenance risk evaluation. Additionally, our results show that language binding and templating tools (such as JSII, WebAssembly, and Protocol Buffers) are widely adopted and in demand. New tools supporting more language targets could significantly expand the cross-ecosystem package landscape.

For future work, researchers can use our taxonomy to investigate trade-offs between native multi-language implementations and translation-based approaches (bindings, templating, wrappers). Security is another direction: because distribution-only packages introduce inter-ecosystem coupling---a vulnerability in one language implementation may propagate to ecosystems that repackage it---our pattern labels could serve as risk indicators in cross-ecosystem dependency tracking. Longitudinal analyses tracking how architectural patterns evolve over a project's lifetime are also warranted.

\section{Threats to Validity}
\label{sec:threats}

In this section, we discuss potential threats to the validity of our study and the measures taken to mitigate them.

\textbf{External Validity.} Our study relies on GitHub repository URLs declared in package registries; packages without GitHub URLs or with incorrect URLs are excluded. This may limit the generalizability of our findings, as some packages---especially those hosted on alternative platforms or with incomplete metadata---are not represented in our dataset. Consequently, our results may not fully capture the diversity of package development and maintenance practices outside of GitHub or in less-documented ecosystems.
We mitigate this by collecting the dataset ourselves and validating the process to the best of our knowledge. We will also make the data available for replication. 

\textbf{Internal Validity.} The ecosystem source file detection uses file extensions as proxies for language and project type, which may miss unconventional file organizations or atypical naming conventions. To assess sensitivity, we re-ran the heuristic-dependent analysis under two composite variants: a \emph{strict} variant (fewer source extensions, 33 additional exclusion patterns, a two-file minimum per ecosystem, and higher P3 coverage thresholds of 90\%/50\%) and a \emph{lenient} variant (broader source extensions, eight fewer exclusion patterns, and lower P3 thresholds of 70\%/30\%). Results are as follows. \PFour (template generation) is fully stable, shifting $\leq 4\%$ under both variants, confirming that its detection via unambiguous IDL file extensions is robust. \PThree (designated-directory) is stable under the lenient variant ($+0.7\%$), but drops by $-17\%$ under the strict variant; the strict drop is almost entirely explained by the two-file minimum, which removes 22\% of fully-matched packages from the input pool—the P3 rate \emph{within} that pool actually rises from 43.5\% to 46.3\%, indicating no loss of detection quality. \PTwo (distribution-only proxy) is similarly driven by the two-file minimum in the strict variant ($+8.4\%$) yet virtually unchanged under lenient ($+0.3\%$). P5 WASM counts shift 18\% (strict) and 11\% (lenient) as expected from the deliberate threshold changes; general binding is stable under lenient ($+0.8\%$) but shifts 16\% under strict due to additional exclusion folders suppressing binding indicators. Taken together, the results confirm that P4 is robustly operationalized; for P3 and P5 the shifts under the strict variant are traceable to specific, documented parameter choices rather than to inherent instability of the detection algorithm. The multi-repo suffix detection was separately validated by checking that the expected language appears in each repository's GitHub-reported language proportions (match rate $> 98\%$), providing an independent cross-check on the heuristic accuracy.

\textbf{Construct Validity.} We use star count as the primary community visibility metric. While star count is widely used in empirical software engineering research~\cite{zerouali-arxiv-2019}, it does not directly measure actual usage, download counts, or dependency adoption. As noted by Zerouali et al.~\cite{zerouali-arxiv-2019}, different methods for measuring popularity can yield different results, particularly in ecosystems like NPM. Thus, our findings regarding package visibility may be influenced by the limitations of this metric.

\textbf{Reliability.} The manual inspection of 200 mismatched packages provides qualitative insights into the causes of mismatches. However, these findings may not generalize to all mismatched cases, as the sample size is limited and subject to inspector bias. Further automated or large-scale manual analyses would be needed to confirm the broader applicability of these observations.

\section{Conclusion}
\label{sec:conclusion}

We present a large-scale empirical study of cross-ecosystem packages across six major package ecosystems.
We find that cross-ecosystem packages account for a small but significant fraction of all GitHub repositories and identify five distinct architectural patterns: multi-repository separation, distribution-only publication, language-specific folder naming, protocol-buffer-like template code generation, and binding/wrapper integration.
Our analysis reveals that distribution-only packages (dominated by Maven WebJar/mvnpm) constitute the largest category and are surprisingly popular, raising concerns that adopters may be unaware of the absence of native source code. Packages using tighter integration strategies (protocol buffers and bindings) tend to be the most actively developed and maintained.


\section*{Declarations}

\subsection*{Funding}
This research was supported by the Japan Society for the Promotion of Science, Grant No. [JP24H00692,JP25K03102,JP26K02889,JP26H02500].

\subsection*{Author Contributions}
Xiangxi Li implemented the code and analysis scripts used in the study, conducted the experiments, and contributed to the writing of the manuscript. Olivier Nourry contributed to the study design, supervised the research, and contributed to the writing of the manuscript. Yoshiki Higo acquired the funding for the project and contributed to the writing of the manuscript. Raula Gaikovina Kula contributed to the study design, supervised the research, and contributed to the writing of the manuscript. All authors read and approved the final manuscript.

\subsection*{Data Availability Statement}
We make all datasets used in this paper publicly available. We also provide all scripts needed to replicate the data mining, filtering, and analysis processes. Upon acceptance, we will provide all raw data for replication.  The scripts can be found at the following link: \href{https://anonymous.4open.science/r/cross-ecosystem-replication-714F}{cross-ecosystem-replication}.

\subsection*{Conflict of Interest}
The authors declare that Raula Gaikovina Kula, is a member of the EMSE Editorial Board. 
All co-authors have seen and agree with the contents of the manuscript and there is no financial interest to report.

\subsection*{Ethical Approval}
Not applicable. This study used publicly available data and did not involve human participants or animals.

\subsection*{Informed Consent}
Not applicable. This study did not involve human participants.

\bibliographystyle{spbasic}      
\typeout{}
\bibliography{bibliography}   


 \vspace{3\baselineskip}

{\setlength\intextsep{0pt}
\begin{wrapfigure}{l}{25mm} 
\includegraphics[width=1in,height=1.25in,clip,keepaspectratio]{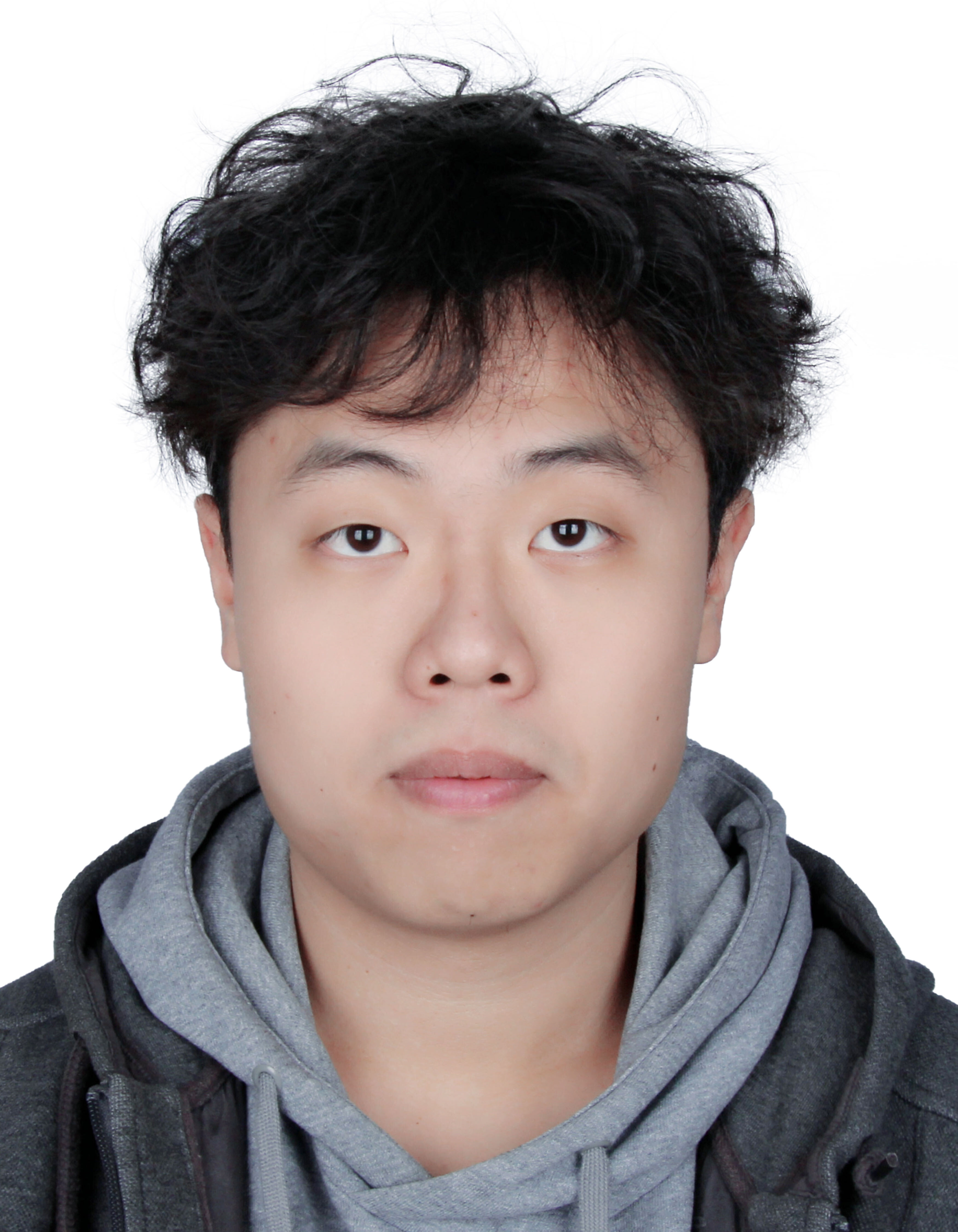}
\end{wrapfigure}\par
\noindent\textbf{Xiangxi Li}\\ Xiangxi Li is bachelor student from China. He is currently on a research exchange program working in Higo Laboratory in Japan.\url{}\par}

\vspace{3\baselineskip}

{\setlength\intextsep{0pt}
\begin{wrapfigure}{l}{25mm} 
    \includegraphics[width=1in,height=1.25in,clip,keepaspectratio]{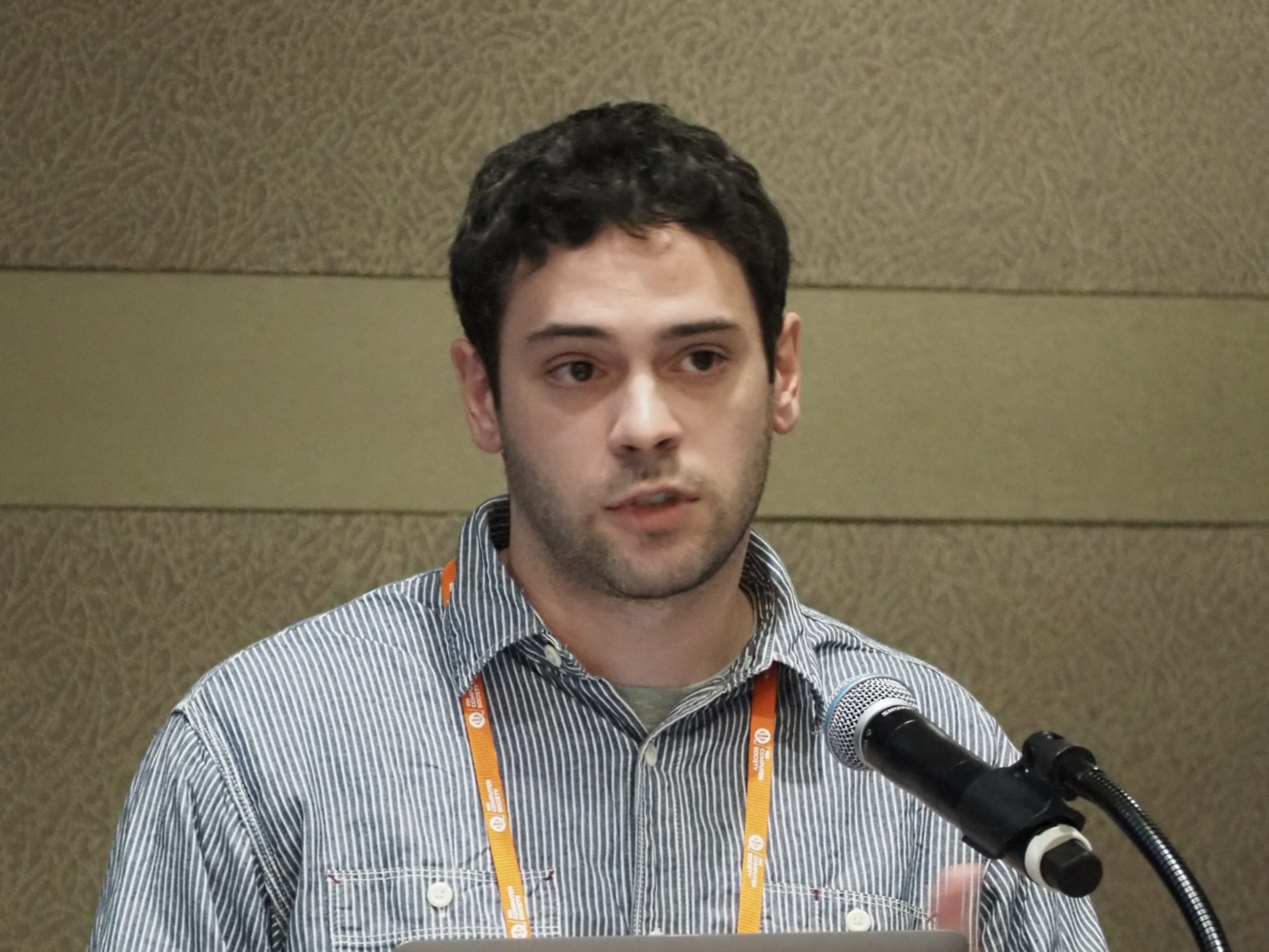}
\end{wrapfigure}\par
\noindent\textbf{Olivier Nourry} is an Assistant Professor in the School of Engineering Science at The University of Osaka. His research interests include empirical software engineering, software maintenance, software quality, software security, and the application of artificial intelligence to software engineering.\url{https://onourry.github.io/olivier-nourry/}. \par}

\vspace{2\baselineskip}

{\setlength\intextsep{0pt}
\begin{wrapfigure}{l}{25mm} 
    \includegraphics[width=1in,height=1.25in,clip,keepaspectratio]{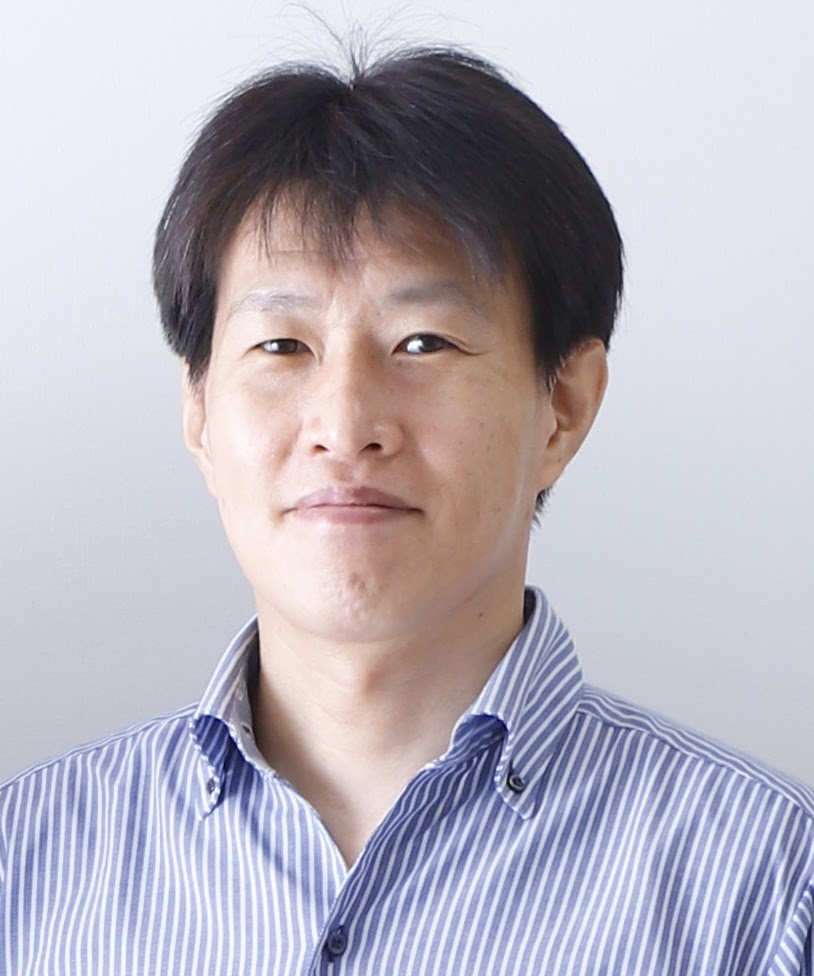}
\end{wrapfigure}\par
\noindent\textbf{Yoshiki Higo}\\ is a Professor in the Graduate School of Information Science and Technology at The University of Osaka. His research interests include software engineering, particularly source code analysis, code clone analysis, refactoring support, software repository mining, and automated program repair. \url{https://sites.google.com/view/yhigo/home}. \par}

\vspace{2\baselineskip}

{\setlength\intextsep{0pt}
\begin{wrapfigure}{l}{25mm} 
    \includegraphics[width=1in,height=1.25in,clip,keepaspectratio]{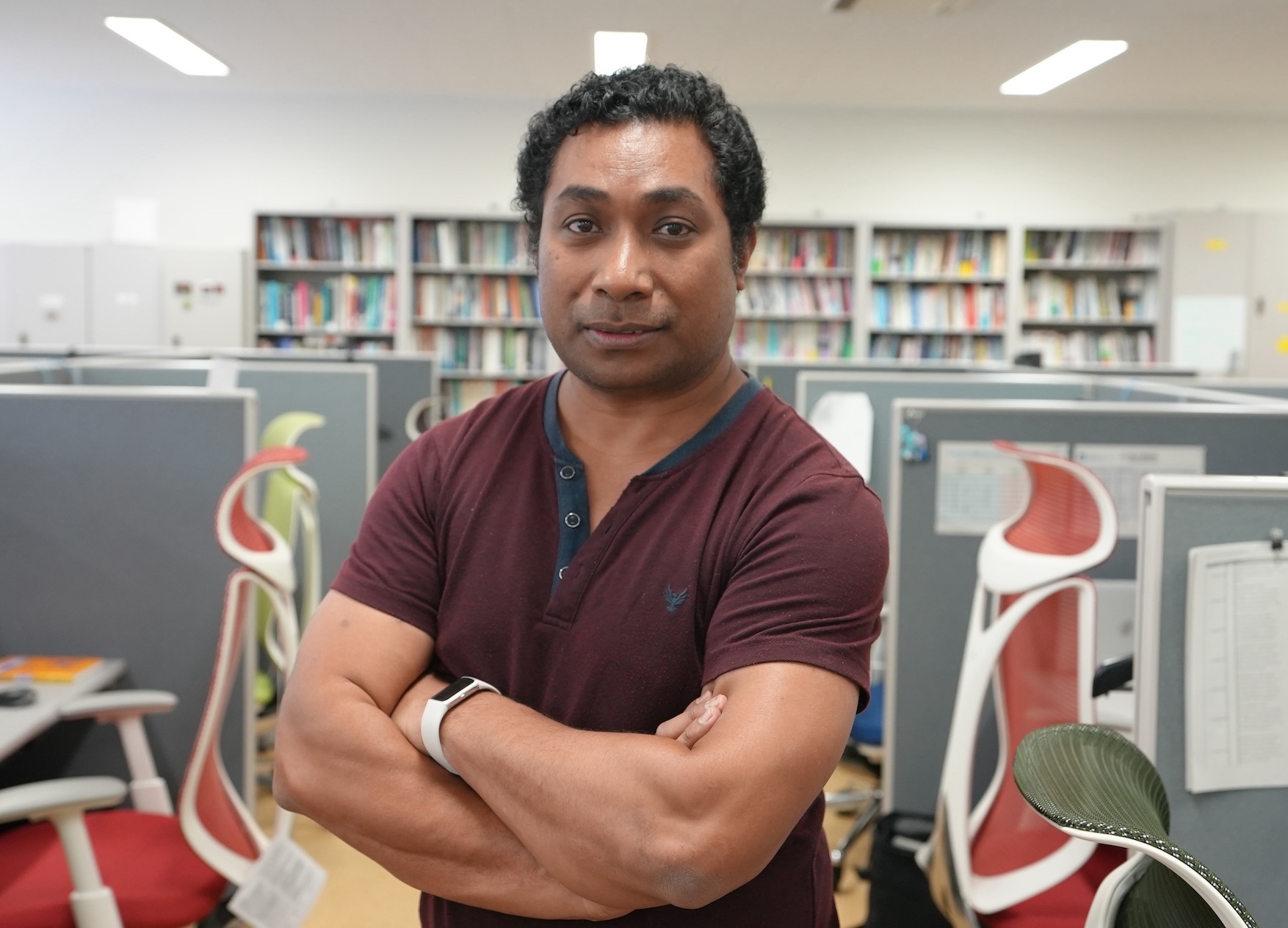}
\end{wrapfigure}\par
\noindent\textbf{Raula Gaikovina Kula}\\ is a Professor at The University of Osaka. He received his Ph.D. degree from NAIST in 2013
and was a Research Assistant Professor at Osaka University. He is active in the
Software Engineering community, serving the community as a PC member for
premium SE venues, some as organizing committee, and reviewer for journals.
His current research interests include library dependencies and security in the
software ecosystem, program analysis such as code clones, and human aspects such as code reviews and coding proficiency. Find him at \url{https://raux.github.io/}
and @augaiko on Twitter. Contact him at raula-k@ist.osaka-u.ac.jp. \par}

\vspace{2\baselineskip}
\end{document}